\documentclass[12pt]{article}
\usepackage[cp1251]{inputenc}

\usepackage{amsfonts,amssymb,amsthm,mathtools}
\usepackage{amsmath}
\usepackage{icomma}
\usepackage{latexsym}
\usepackage{graphicx}
\usepackage{subfigure}
\usepackage {bm}

\usepackage{geometry} 
\begin{document}
	
	\begin{center}
		\textbf{THE FOLDY-WOUTHUYSEN TRANSFORMATION WITH DIRAC MATRICES IN THE 
			CHIRAL REPRESENTATION: CLOSED EXPRESSIONS FOR ENERGY OPERATORS OF FERMIONS MOVING IN ELECTROSTATIC FIELDS}
	\end{center}

	\begin{center}
		
		{V.~P.~Neznamov\footnote{vpneznamov@mail.ru, vpneznamov@vniief.ru}}\\
		
		\hfil
		{\it \mbox{	Russian Federal Nuclear Center--All-Russian Research Institute of Experimental Physics},  Mira pr., 37, Sarov, 607188, Russia} \\
	\end{center}


\begin{abstract}
\noindent
\footnotesize{In this paper, we obtain the closed expressions for energy operators in the Foldy-Wouthuysen representation in the presence of electrostatic fields. We also establish the connection between the Foldy-Wouthuysen representation and the Feynman-Gell-Mann representation.} \\



\end{abstract}


\section{Introduction}	

The paper devoted to the Foldy-Wouthuysen (FW) representation was first 
published in 1950 \cite{bib1}. The transition to the FW representation is implemented by the unitary transformation of the Dirac equation. In Ref. \cite{bib1}, in addition to the case 
of the free motion of an electron, the Hamiltonian $H_{FW} $ was found in the 
non-relativistic approximation in the form of the series in terms of the 
power of $1/m$ for the case of the electron motion in the 
external electromagnetic field.

Blount \cite{bib2} found the Hamiltonian $H_{FW} $ in the 
form of the series in terms of powers of electromagnetic fields and their 
time and space derivatives. Case \cite{bib3} obtained the 
exact FW transformation in the presence of the external magnetostatic field 
${\rm {\bf B}}= \rm {rot} {\rm {\bf A}}$. In Refs. \cite{bib4} and \cite{bib5}, the author examined the quantum 
electrodynamics in the FW representation. In these papers, the relativistic 
Hamiltonian $H_{FW} $ was obtained in the form of the series in terms of 
powers of the charge $e$.

In the paper Ref. \cite{bib6}, the necessary and sufficient 
conditions for the transition to the FW representation is shown. The 
necessary condition is a diagonalization of the Hamiltonian $H_{FW}$ relative to mixing of the upper and lower spinors of the bispinor wave 
function $\psi_{FW} $.

The sufficient condition is the reduction of the wave function $\psi_{FW} 
$ versus the fermion energy sign. For $\varepsilon >0$, the lower spinor of 
the wave function vanishes, for $\varepsilon <0$, the upper spinor vanishes.

Since the publication of the first paper Ref. \cite{bib1}, in 
the literature, the results of numerous studies of the practical application 
of the FW representation have been presented. The development of the quantum 
electrodynamicin the FW representation (QED)$\rm{_{FW}}$ is one of them 
\cite{bib4, bib5} (see also Ref. \cite{bib7}). A (QED)$\rm{_{FW}}$ fermion vacuum is empty 
without the opportunity of the creation and the annihilation of the 
electron-positron pairs. (QED)$\rm{_{FW}}$ in Ref. \cite{bib4} and \cite{bib5} is represented as the perturbation theory series in terms of the powers of the charge $e$.

In the presence of the strong electromagnetic fields (i.e. for the case of 
the nonperturbative (QED)$\rm{_{FW}}$), it is necessary to use the closed 
expressions for the Hamiltonians and the energy operators. There aren't such 
expressions until now.

In the paper, we obtain the closed expressionsfor the energy operators in the 
FW representation. In this case, we use the Dirac matrices in the chiral 
representationin the presence of electrostatic fields.

The closed expressions for the energy operators in the nonperturbative 
(QED)$\rm{_{FW}}$ will be useful in prediction and interpretation of the results 
from the experiments with the strong electrostatic fields either atthe 
collisions of heavy ions with $Z>10$ or when focusing some high intensity 
laser pulses $\left( I > 10^{20}\,\ W/cm^2 \right)$.

We also establish the connection between the FW representation and the 
Feynman-Gell-Mann representation (FGM) \cite{bib8} in the 
presence of electrostatic fields. In FGM representation, one can also 
construct a quantum electrodynamics with the final results which are 
equivalent to the standard QED \cite{bib9}. In particular, a 
value of the FGM representation is a simplicity of transition to the 
electromagnetic fields depending on time (see Eqs. (\ref{eq47}), (\ref{eq49}) and (\ref{eq50})).The 
further studies will show an opportunity (impossibility) of such transition 
in the FW representation.

This paper is arranged as follows. Section 2 contains the relationships associated with the use of the 
standard and chiral representations of the Dirac matrices. In Sec. 3, the 
FW transformation with Dirac matrices in the chiral 
representation(CFW transformation) is performed. Section 4 presents the 
FGM transformation and the discussion of the relation 
between this representation and the FW representation. The results of our studies are formulated in conclusion section (Sec.5).

\section{The Dirac equation in external static electromagnetic fields}
\label{Sec2}

The Dirac equation in the external electromagnetic field can be written as
\begin{equation}
	\label{eq1}
	p^{0}\psi_{D} ({{\bf x}},t)=\left( {{\rm {\bm {\alpha \pi} }}+\beta 
		m+eA^{0}({\rm {\bf x}},t)} \right)\psi_{D} ({\rm {\bf x}},t).
\end{equation}
Hereinafter, the system of units of $\hslash =c=1$ is used; $\psi_{D} ({\rm 
	{\bf x}},t)$ is the bispinor wave function; $p^{\mu }=i\dfrac{\partial 
}{\partial x_{\mu } };\,\,\,\mu =0,1,2,3$; ${\rm {\bm \pi }}({\rm {\bf 
		x}},t)={\rm {\bf p}}-e{\rm {\bf A}}({\rm {\bf x}},t)$; $A^{\mu }({\rm {\bf 
		x}},t)$ is electromagnetic 4-potential; $\alpha^{i},\,\,\beta $ are 
four-dimensional Dirac matrices $\left( {i=1,2,3} \right)$; in the standard 
representation of matrices $\alpha^{i},\,\,\beta ,\,\,\Sigma^{i},\,\,\gamma^{5},\,\,\gamma^{i}$ have the following form:
\begin{equation}
	\label{eq2}
	\begin{array}{l}
		\alpha^{i}=\left( {\begin{array}{l}
			0\;\;\,\sigma^{i}{\kern 1pt} \\ 
			\sigma^{i}\;\;0 \\ 
	\end{array}} \right),\,\,\beta =\gamma^{0}=\left( {\begin{array}{l}
			I\;\;\,{\kern 5pt}0 \\ 
			0\;\,-I \\ 
				\end{array}} \right),\,\,\Sigma^{i} =\left( {\begin{array}{l}
			\sigma^{i}\;\;\,{\kern 1pt}0 \\ 
			0\;\,\,\,\,\sigma^{i} \\ 
	\end{array}} \right), \\ [15pt]
\gamma^{5} =\left( {\begin{array}{l}
			0\;\;\,{\kern 1pt}I \\ 
			I\;\;\;0 \\ 
	\end{array}} \right),\,\,\gamma^{i}=\gamma^{0}\alpha^{i}.
\end{array}
\end{equation}
Sometimes, it is convenient to use the chiral representation of the Dirac 
matrices
\begin{equation}
	\label{eq3}
	\begin{array}{l}
	\alpha^{i}_{c}=S \alpha^{i} S^{-1} = \left( {\begin{array}{l}
			\sigma^{i}\;\;\,{\kern 5pt} 0{\kern 1pt} \\ 
			0\;\;-\;\sigma^{i} \\ 
	\end{array}} \right),\;\,\beta_{c} =\gamma^{0}_{c}= S \beta S^{-1} =\left( {\begin{array}{l}
			0\;\;\,{\kern 1pt}I \\ 
			I\;\,\,0 \\ 
	\end{array}} \right), \\ [15pt]
    \Sigma^{i}_{c} = S \Sigma^{1} S^{-1} =  \left( {\begin{array}{l}
    		\sigma^{i}\;\;\,0{\kern 1pt} \\ 
    		0\;\;\;\sigma^{i} \\ 
    \end{array}} \right),   
    \gamma^{5}_{c} = S \gamma^{5} S^{-1} = \left( {\begin{array}{l}
			I\;\;\,\,\,{\kern 5pt}0 \\ 
			0\;-I \\ 
	\end{array}} \right),\;\gamma^{i}_{c}=\gamma^{0}_{c}\alpha^{i}_{c}.
	\end{array}
\end{equation}

In (\ref{eq3}), the matrix of unitary transformation $S$ has the following form:
\[
S=S^{+}=S^{-1}=\frac{1}{\sqrt{2}} \left( {\begin{array}{l}
		I\;\;\,{\kern 5pt}I \\ 
		I\;-I \\ 
\end{array}} \right).
\]

In the standard representation (\ref{eq2}), the matrices $\beta,\,\Sigma^{i}$ are even (they do not mix up the upper and lower components of the wave function). The matrices $\alpha^{i},\,\gamma^{5},\,\gamma^{i}$ are odd (they mix up the upper and lower components of the wave function).

In the chiral representation (\ref{eq3}), matrices $\alpha^{i}_{c},\,\Sigma^{i}_{c},\,\gamma^{5}_{c}$ are even. Matrices $\beta_{c},\,\gamma_{c}^{i}$ are odd.

The diagonal matrices $\Sigma^{i},\,\Sigma_{c}^{i}$ are even. The matrices $\gamma^{i},\,\gamma^{i}_{c}$ are odd. The helicity operators 

\[
\Pi^{\pm}=\Pi^{\pm}_{c}=\frac{1}{2} \left( {\begin{array}{l}
		I \pm \dfrac{\bm \sigma \bf p}{\left| \bf p \right| }\;\;\,{\kern 1pt}\,\,\,\,0 \\
		\,\,\,\,\,\,\,0\;\;\;\;\;\;\;\;\; I \pm \dfrac{\bm \sigma \bf p}{\left| \bf p \right| } \\ 
\end{array}} \right)
\] are even.

The FW transformations always eliminate terms with the odd matrices.

The Chiral representation (\ref{eq3}) is widely used in modern gauge field theories 
and in the Standard Model, in particular.

It what follows, we use representation (\ref{eq3}) due to the opportunity to obtain closed 
expressions for energy operators after FW transformations in 
the presence of anelectrostatic field.

Let us consider static external fields $A^{i}({\rm {\bf x}}),\,A^{0}({\rm 
	{\bf x}})$. In this case, the stationary states are implemented
\begin{equation}
	\label{eq4}
	\psi_{D} ({\rm {\bf x}},t)=e^{-i\varepsilon t}\psi_{D} ({\rm {\bf x}}).
\end{equation}
For the standard representation of the Dirac matrices (\ref{eq2}), Eq. (\ref{eq1}) can be written as
\begin{equation}
	\label{eq5}
	\begin{array}{l}
		\varepsilon \varphi \left( {{\rm {\bf x}}} \right)={\rm {\bm {\sigma \pi }
		}}\chi \left( {{\rm {\bf x}}} \right)+m\,\varphi \left( {{\rm {\bf x}}} 
		\right)+eA^{0}\varphi \left( {{\rm {\bf x}}} \right), \\ [5pt]
		\varepsilon \chi \left( {{\rm {\bf x}}} \right)={\rm {\bm {\sigma \pi }
		}}\varphi \left( {{\rm {\bf x}}} \right)-m\,\chi \left( {{\rm {\bf x}}} 
		\right)+eA^{0}\chi \left( {{\rm {\bf x}}} \right), \\ [5pt]
		\chi \left( {{\rm {\bf x}}} \right)=\dfrac{1}{\varepsilon +m-eA^{0}}{\rm 
			{\bm \sigma }}\left( {{\rm {\bf p}}-e{\rm {\bf A}}} \right)\varphi \left( 
		{{\rm {\bf x}}} \right), \\ [5pt]
		\varphi \left( {{\rm {\bf x}}} \right)=\dfrac{1}{\varepsilon -m-eA^{0}}{\rm 
			{\bm \sigma }}\left( {{\rm {\bf p}}-e{\rm {\bf A}}} \right)\chi \left( {{\rm 
				{\bf x}}} \right). \\ 
	\end{array}
\end{equation}
In (\ref{eq5}), the bispinor $\psi_{D} ({\rm {\bf x}})$ is represented as a column of two spinors $\varphi \left( {{\rm {\bf x}}} \right),\,\chi \left( {{\rm 
		{\bf x}}} \right)$
\begin{equation}
	\label{eq6}
	\psi_{D} ({\rm {\bf x}})=\left( {{\begin{array}{*{20}c}
				{\varphi ({\rm {\bf x}})} \hfill \\
				{\chi ({\rm {\bf x}})} \hfill \\
	\end{array} }} \right).
\end{equation}
In (\ref{eq5}), the operator of momentum ${\rm {\bf p}}$ does not commute 
with $eA^{0}\left( {{\rm {\bf x}}} \right)$. The energy $\varepsilon$ in (\ref{eq4}) and (\ref{eq5}) is a {\it{c}}-number.

The similar relationships for the chiral representation of the Dirac 
matrices have the form
\begin{equation}
	\label{eq7}
	\begin{array}{l}
		\varepsilon \varphi_{c} \left( {{\rm {\bf x}}} \right)={\rm {\bm {\sigma 
				\pi} }}\varphi_{c} \left( {{\rm {\bf x}}} \right)+m\,\chi_{c} \left( {{\rm 
				{\bf x}}} \right)+eA^{0}\varphi_{c} \left( {{\rm {\bf x}}} \right), \\ [5pt]
		\varepsilon \chi_{c} \left( {{\rm {\bf x}}} \right)=-{\rm {\bm {\sigma \pi }
		}}\chi_{c} \left( {{\rm {\bf x}}} \right)+m\,\varphi_{c} \left( {{\rm 
				{\bf x}}} \right)+eA^{0}\chi_{c} \left( {{\rm {\bf x}}} \right), \\ [5pt] 
		\chi_{c} \left( {{\rm {\bf x}}} \right)=\dfrac{1}{\varepsilon +{\rm {\bm 
					{\sigma \pi} }}-eA^{0}}m\varphi_{c} \left( {{\rm {\bf x}}} \right), \\ [10pt]
		\varphi_{c} \left( {{\rm {\bf x}}} \right)=\dfrac{1}{\varepsilon -{{\bm 
				{	\sigma \pi} }}-eA^{0}}m\chi_{c} \left( {{\rm {\bf x}}} \right), \\ 
	\end{array}
\end{equation}
\begin{equation}
	\label{eq8}
	\psi_{D} ({\rm {\bf x}})=\left( {{\begin{array}{*{20}c}
				{\varphi_{c} ({\rm {\bf x}})} \hfill \\
				{\chi_{c} ({\rm {\bf x}})} \hfill \\
	\end{array} }} \right).
\end{equation}

The relationships (\ref{eq5}) and (\ref{eq7}) turn into each other at $m \leftrightarrow \bm {\sigma \pi}$; the matrices (\ref{eq2}) and (\ref{eq3}) turn into each other at $\beta \leftrightarrow \gamma_{c}^{5}$.

In the relationships (\ref{eq5}), for determining $\,\chi \left( {{\rm {\bf x}}} 
\right),\,\varphi \left( {{\rm {\bf x}}} \right)$, the numerators of the 
expressions do not commute with the appropriate denominators. In the 
relationships (\ref{eq7}), for determining $\chi_{c} \left( {{\rm {\bf x}}} 
\right),\,\varphi_{c} \left( {{\rm {\bf x}}} \right)$, the numerators of 
the expressions commute with the appropriate denominators. It allows to 
obtaining closed expressions for the energy operators in the 
FW representation with electrostatic fields.

Our method is applicable to all gauge-invariant electromagmetic fields interacting with the Dirac particles according to the Eq. (\ref{eq1}). In this case, the introduction of electromagnetic interactions in the Dirac equation is implemented by replacing $p^{\mu} \to p^{\mu} - eA^{\mu} \left( {\bf x}, t \right) $ .

If the numerators in (\ref{eq7}) do not commute with the appropriate denominators, 
the transition to the FW representation by means of the proposed 
method becomes impossible. In particular, this situation refers to a motion 
of the Dirac neutron with a magnetic moment $\mu$ in electrostatic 
field ${\rm {\bf E}}$ \cite{bib10}. For this case Eriksen 
implemented the transition to FW representation by means of his own method 
with the use of the Dirac matrices in the standard representation \cite{bib11}.

\section{The FW transformation with Dirac matrices in the chiral representation}
\label{Sec3}

In the FW representation, the two conditions must be fulfilled 
(see, for instance, Ref. \cite{bib6}):

\begin{itemize}
	\item[(i)] The Hamiltonian or the energy operator are diagonal relative to the upper and lower spinors of the transformed wave function $\psi_{FW} ({\rm {\bf x}})$, i.e., these operators do not mix up the upper and lower components $\psi_{FW} ({\rm {\bf x}})$.
	\item[(ii)] At the FW transformation, the condition of the wave function reduction must be met. For the case when the Dirac Hamiltonian does not depend on time (the case of static external fields), the condition of reduction can be written as \footnote{In what follows, the wave functions are normalized by unitary probability in the box of volume $V$. For brevity sake, in our expressions for wave functions, there are no multipliers $1 / {\sqrt V }$.}
\end{itemize}
\begin{equation}
	\label{eq10}
	\begin{array}{l}
		\psi_{D}^{\left( + \right)} ({\rm {\bf x}},t)=e^{-i\varepsilon t}A_{\left( 
		+ \right)} \left( {{\begin{array}{*{20}c}
				{\varphi_{c} ({\rm {\bf x}})} \hfill \\
				{\chi_{c} ({\rm {\bf x}})} \hfill \\
	\end{array} }} \right)\to \psi_{FW}^{\left( + \right)} ({\rm {\bf 
			x}},t)=U_{FW}^{\left( + \right)} ({\rm {\bf x}})\psi_{D}^{\left( + \right)} 
	({\rm {\bf x}},t)= \\
	= e^{-i\varepsilon t}\left( {{\begin{array}{*{20}c}
				{\varphi_{c} ({\rm {\bf x}})} \hfill \\
				0 \hfill \\
	\end{array} }} \right),
\end{array}
\end{equation}
where $\varepsilon >0$.
\begin{equation}
	\label{eq11}
	\begin{array}{l}
		\psi_{D}^{\left( - \right)} ({\rm {\bf x}},t)=e^{-i\varepsilon t}A_{\left( 
		- \right)} \left( {{\begin{array}{*{20}c}
				{\varphi_{c} ({\rm {\bf x}})} \hfill \\
				{\chi_{c} ({\rm {\bf x}})} \hfill \\
	\end{array} }} \right)\to \psi_{FW}^{\left( - \right)} ({\rm {\bf 
			x}},t)=U_{FW}^{\left( - \right)} ({\rm {\bf x}})\psi_{D}^{\left( - \right)} 
	({\rm {\bf x}},t)= \\
	= e^{-i\varepsilon t}\left( {{\begin{array}{*{20}c}
				0 \hfill \\
				{\chi_{c} ({\rm {\bf x}})} \hfill \\
	\end{array} }} \right),
	\end{array}
\end{equation}
where $\varepsilon <0$.

In (\ref{eq10}) and (\ref{eq11}), $A_{\left( + \right)}$ and $A_{\left( - \right)} $ are normalization operators. $U_{FW}^{\left( + \right)}$ and $U_{FW}^{\left( - 
	\right)} $ are operators of the CFW transformation. Operators $A_{\left( + 
	\right)} ,\,\,A_{\left( - \right)}$ and $U_{FW}^{\left( + \right)} 
,\,\,\,U_{FW}^{\left( - \right)} $ are not necessarily identical for 
positive and negative energies.

The operators of transformations $U_{FW}^{\left( + \right)} 
,\,\,\,U_{FW}^{\left( - \right)} $ are determined, as a rule, by the 
fulfillment of condition (i). Thereafter, the fulfillment of condition (ii) is 
checked \cite{bib6}.

In this paper, the type of the operators $U_{FW}^{\left( + \right)} 
,\,\,\,U_{FW}^{\left( - \right)} $ is determined from the reduction 
condition of the wave function (see (\ref{eq10}), (\ref{eq11})). Thereafter, the energy 
operator is determined by the fulfillment of condition (i).

First, let us consider the case of positive energies $\varepsilon =\left| E 
\right|>0$.

From Eqs. (\ref{eq7}) and (\ref{eq8}), the wave function $\psi_{D}^{\left( + \right)} 
({\rm {\bf x}})$ can be written as
\begin{equation}
	\label{eq12}
	\psi_{D}^{\left( + \right)} ({\rm {\bf x}})=A_{\left( + \right)} \left( 
	{{\begin{array}{*{20}c}
				{\kern 45pt}{\varphi_{c} ({\rm {\bf x}})} \hfill \\ [10pt]
				{\dfrac{1}{\left| E \right|+{\rm {\bm {\sigma \pi} }}-eA^{0}\left( {{\rm {\bf 
									x}}} \right)}m\varphi_{c} ({\rm {\bf x}})} \hfill \\
	\end{array} }} \right).
\end{equation}
The normalization factor $A_{\left( + \right)} $ can be found from the 
condition
\begin{equation}
	\label{eq13}
	\psi_{D}^{\left( + \right)} ({\rm {\bf x}})^{\dag }\psi_{D}^{\left( + 
		\right)} ({\rm {\bf x}})=\varphi_{c}^{\dag } ({\rm {\bf x}})\varphi_{c} 
	({\rm {\bf x}}).
\end{equation}
In (\ref{eq13}) and in what follows, the sign ''$\dag $'' stands for ''Hermitian conjugation''.

Then, from (\ref{eq12}) and (\ref{eq13}),
\begin{equation}
	\label{eq14}
	\begin{array}{c}
		A_{\left( + \right)}^{2} \left( {1+\dfrac{m^{2}}{\left( {\left| E 
				\right|+{\rm {\bm {\sigma \pi} }}-eA^{0}\left( {{\rm {\bf x}}} \right)} 
			\right)^{2}}} \right)=1, \\ [15pt]
		A_{\left( + \right)} =\left( 
	{1+\dfrac{m^{2}}{\left( {\left| E \right|+{\rm {\bm {\sigma \pi} 
				}}-eA^{0}\left( {{\rm {\bf x}}} \right)} \right)^{2}}} \right)^{-1 
		\mathord{\left/ {\vphantom {1 2}} \right. \kern-\nulldelimiterspace} 2}.
	\end{array}
\end{equation}
Hence, we can easily determine the view of the unitary 
operator $U_{FW}^{\left( + \right)\dag } =\,\left( {U_{FW}^{\left( + \right)} 
} \right)^{-1}$
\begin{equation}
	\label{eq15}
	U_{FW}^{\left( + \right)} =A_{\left( + \right)} \left( {1+\frac{1}{\left| E 
			\right|+{\rm {\bm {\sigma \pi} }}-eA^{0}\left( {{\rm {\bf x}}} \right)}\gamma 
		^{5}_{c} \beta_{c} m} \right).
\end{equation}
The wave function in the FW representation has the following form:
\begin{equation}
	\label{eq16}
	\begin{array}{l}
		\psi_{FW}^{\left( + \right)} ({\rm {\bf x}})=U_{FW}^{\left( + \right)} 
		\psi_{D}^{\left( + \right)} ({\rm {\bf x}})=A_{\left( + \right)} \left( 
		{1+\dfrac{1}{\left| E \right|+{\rm {\bm {\sigma \pi} }}-eA^{0}\left( {{\rm {\bf 
							x}}} \right)}\gamma^{5}_{c} \beta_{c} m} \right)\times \\ [10pt]
		\times A_{\left( + \right)} \left( {{\begin{array}{*{20}c}
				\,\,\,\,\,\,\,\,\,\,\,\,\,\,\,\,\,\,\,\,\,\,\,\,\,\,\,\,{\varphi_{c} ({\rm {\bf x}})} \hfill \\
					{\dfrac{1}{\left| E \right|+{\rm {\bm {\sigma \pi} }}-eA^{0}\left( {{\rm {\bf 
										x}}} \right)}m\varphi_{c} ({\rm {\bf x}})} \hfill \\
		\end{array} }} \right)=\left( {{\begin{array}{*{20}c}
					{\varphi_{c} ({\rm {\bf x}})} \hfill \\
					\,\,\,\,\,\,0 \hfill \\
		\end{array} }} \right). \\ 
	\end{array}
\end{equation}
Let us consider now the case of negative energies $\varepsilon =-\left| E 
\right|<0$. From Eqs. (\ref{eq7}) and (\ref{eq8}), the wave function $\psi_{D}^{\left( 
	- \right)} ({\rm {\bf x}})$ can be written as
\begin{equation}
	\label{eq17}
	\psi_{D}^{\left( - \right)} ({\rm {\bf x}})=A_{\left( - \right)} \left( 
	{{\begin{array}{*{20}c}
				{-\dfrac{1}{\left| E \right|+{\rm {\bm {\sigma \pi }}}+eA^{0}\left( {{\rm {\bf 
									x}}} \right)}m\chi_{c} ({\rm {\bf x}})} \hfill \\ [10pt]
				\,\,\,\,\,\,\,\,\,\,\,\,\,\,\,\,\,\,\,\,\,\,\,\,\,\,\,\,\,\,{\chi_{c} ({\rm {\bf x}})} \hfill \\
	\end{array} }} \right).
\end{equation}
Here,
\begin{equation}
	\label{eq18}
	A_{\left( - \right)} =\left( {1+\dfrac{m^{2}}{\left( {\left| E 
				\right|+{\rm {\bm {\sigma \pi} }}+eA^{0}\left( {{\rm {\bf x}}} \right)} 
			\right)^{2}}} \right)^{-1 \mathord{\left/ {\vphantom {1 2}} \right. 
			\kern-\nulldelimiterspace} 2}.
\end{equation}
The unitary operator $U_{FW}^{\left( - \right)} $ is
\begin{equation}
	\label{eq19}
	U_{FW}^{\left( - \right)} =A_{\left( - \right)} \left( {1+\dfrac{1}{\left| E 
			\right|+{\rm {\bm {\sigma \pi} }}+eA^{0}\left( {{\rm {\bf x}}} \right)}\gamma 
		^{5}_{c} \beta_{c} m} \right).
\end{equation}
The wave function in the Foldy-Wouthuysen representation for negative 
energies $\varepsilon $ have the following form:
\begin{equation}
	\label{eq20}
	\begin{array}{l}
		\psi_{FW}^{\left( - \right)} ({\rm {\bf x}})=U_{FW}^{\left( - \right)} 
		\psi_{D}^{\left( - \right)} ({\rm {\bf x}})=A_{\left( - \right)} \left( 
		{\dfrac{1}{\left| E \right|+{\rm {\bm {\sigma \pi} }}+eA^{0}\left( {{\rm {\bf 
							x}}} \right)}\gamma^{5}_{c} \beta_{c} m} \right)\times \\ [10pt]
		\times A_{\left( - \right)} \left( {{\begin{array}{*{20}c}
					{-\dfrac{1}{\left| E \right|+{\rm {\bm {\sigma \pi} }}+eA^{0}\left( {{\rm {\bf 
										x}}} \right)}m\chi_{c} ({\rm {\bf x}})} \hfill \\
					\,\,\,\,\,\,\,\,\,\,\,\,\,\,\,\,\,\,\,\,\,\,\,\,\,\,\,\,\,\,\,{\chi_{c} ({\rm {\bf x}})} \hfill \\
		\end{array} }} \right)=\left( {{\begin{array}{*{20}c}
					\,\,\,\,\,0 \hfill \\
					{\chi_{c} ({\rm {\bf x}})} \hfill \\
		\end{array} }} \right). \\ 
	\end{array}
\end{equation}
Equalities (\ref{eq16}) and (\ref{eq20}) show the fulfillment of the reduction condition of the 
wave function when performing transformations by the operators 
$U_{FW}^{\left( + \right)}$ and $U_{FW}^{\left( - \right)} $.

Now, let us write the expressions for the Hamiltonians in the 
FW representation
\begin{equation}
	\label{eq21}
	H_{FW}^{( + )} =U_{FW}^{( + )} H_{D} U_{FW}^{(
+)\dag},
\end{equation}
\begin{equation}
	\label{eq22}
	H_{FW}^{\left( - \right)} =U_{FW}^{\left( - \right)} H_{D} U_{FW}^{\left( 
		- \right) \dag}.
\end{equation}
$H_{FW}^{\left( + \right)} $ acts on function (\ref{eq16}); in this case,
\begin{equation}
	\label{eq23}
	\gamma^{5}_{c} \psi_{FW}^{\left( + \right)} =\psi_{FW}^{\left( + \right)} .
\end{equation}
$H_{FW}^{\left( - \right)} $ acts on function (\ref{eq20}); in this case,
\begin{equation}
	\label{eq24}
	\gamma^{5}_{c} \psi_{FW}^{\left( - \right)} =-\psi_{FW}^{\left( - \right)} .
\end{equation}


\subsection{Hamiltonians and FW energy operators for positive $\varepsilon =\left| E \right|>0$}
\label{Subsec3.1}

So,
\begin{equation}
	\label{eq25}
	H_{FW}^{\left( + \right)} =U_{FW}^{\left( + \right)} H_{D} U_{FW}^{\left( 
		+ \right) \dag }=H_{even}^{\left( + \right)} +H_{odd}^{\left( + \right)}.
\end{equation}
Here, $H_{odd}^{\left( + \right)} $ is a part of the Hamiltonian with the 
matrix $\beta_{c} $, mixing up the upper and lower components of the wave 
function.

In the FW representation, according to condition (i)
\begin{equation}
	\label{eq26}
	H_{odd}^{\left( + \right)} \psi_{FW}^{\left( + \right)} =0.
\end{equation}
When performing (\ref{eq26}), the Hamiltonian $H_{FW}^{\left( + \right)} 
=H_{even}^{\left( + \right)} $ does not contain summands with the matrix 
$\beta_{c} $.

Equation (\ref{eq26}) conditions the operator view of energy $\left| E \right|$.

Let us determine the view of the operator $H_{odd}^{\left( + \right)} $ from 
Eqs. (\ref{eq1}), (\ref{eq3}) and (\ref{eq15}), (\ref{eq25})
\begin{equation}
	\label{eq27}
	\begin{array}{l}
		H_{odd}^{\left( + \right)} =\beta_{c} mA_{\left( + \right)} \left[ 
		{1-\dfrac{m^{2}}{\left( {\left| E \right|+{\rm {\bm {\sigma \pi} }}-eA^{0}} 
				\right)^{2}}}  - 
		\dfrac{1}{\left| E \right|+{\rm {\bm {\sigma \pi} }}-eA^{0}}\left( 
			{{\rm {\bm {\sigma \pi} }}+\gamma^{5}_{c} eA^{0}} \right)- \right. \\ [15pt]
		\left. 	{-\left( {{\rm {\bm {\sigma \pi} }}-\gamma^{5}_{c} eA^{0}} 
			\right)\dfrac{1}{\left| E \right|+{\rm {\bm {\sigma \pi} }}-eA^{0}}} 
		\right]A_{\left( + \right)} . \\ 
	\end{array}
\end{equation}
Taking into account Eq. (\ref{eq23}), the numerator and the denominator of the 
last summand in (\ref{eq27}) commute with each other. Then,
\begin{equation}
	\label{eq28}
	H_{odd}^{\left( + \right)} =\beta_{c} mA_{\left( + \right)} \left[ 
	{1-\frac{m^{2}}{\left( {\left| E \right|+{\rm {\bm {\sigma \pi} }}-eA^{0}} 
			\right)^{2}}-\frac{2}{\left| E \right|+{\rm {\bm {\sigma \pi} }}-eA^{0}}{\rm 
			{\bm {\sigma \pi} }}} \right]A_{\left( + \right)} .
\end{equation}
Taking into account (\ref{eq28}), the Eq. (\ref{eq26}) allows writing the following expression
\begin{equation}
	\label{eq29}
	\left( {\left( {\left| E \right|-eA^{0}} \right)^{2}-\left( {{\rm {\bm 
					{\sigma \pi} }}} \right)^{2}-m^{2}+ie{\rm {\bm \sigma }}\nabla A^{0}} 
	\right)\varphi_{FG} \left( {{\rm {\bf x}}} \right)=0.
\end{equation}
Here, $\varphi_{FG} \left( {{\rm {\bf x}}} \right)$ is the spinor wave 
function in the FGM representation (see the following section 
of the paper)
\begin{equation}
	\label{eq30}
	\varphi_{FG} \left( {{\rm {\bf x}}} \right)=A_{\left( + \right)} \varphi 
	_{c} \left( {{\rm {\bf x}}} \right).
\end{equation}
Theoretically, we can perform transformation of the similarity in Eq. 
(\ref{eq30}) and obtain an equation with the use of only a spinor wave function in 
the FW representation
\begin{equation}
	\label{eq31}
	\varphi_{c} =A_{\left( + \right)}^{-1} \varphi_{FG} ,
\end{equation}
\begin{equation}
	\label{eq32}
	\left[ {A_{\left( + \right)}^{-1} \left( {\left( {\left| E \right|-eA^{0}} 
			\right)^{2}-\left( {{\rm {\bm {\sigma \pi} }}} \right)^{2}-m^{2}+ie{\rm {\bm 
					\sigma }}\nabla A^{0}} \right)A_{\left( + \right)} } \right]\varphi 
	_{c} =0.
\end{equation}
Transformations (\ref{eq31}) and (\ref{eq32}) can be implemented, taking into account the 
view of the operator $A_{\left( + \right)} $ (see formula (\ref{eq14})). However, the 
practical realization of these transformations is difficult because of 
absence of commutation of the appropriate operators.

We know that the similarity transformation preserves the energy spectrum of 
a transformed equation. Therefore, in our case, the spectra of Eqs. 
(\ref{eq32}) and (\ref{eq29}) are identical. If we want to determine the energy of 
stationary states in the FW representation (Eq. (\ref{eq32})), we can analyze simpler Eq. (\ref{eq29}) with the wave function in the 
FGM representation.

Let us consider some extreme cases.
\begin{itemize}
	\item[(i)] $A^{0} \left( {{\rm {\bf x}}} \right)=A^{i} \left( {{\rm {\bf x}}} \right)=0$ (free motion).
	
In this case, the operator $A_{\left( + \right)} $ in (\ref{eq32}) commutes with the summands on the left and the equation becomes
\begin{equation}
	\label{eq33}
	\left( {\left| E \right|^{2}-m^{2}-{\rm {\bf p}}^{2}} \right)\varphi_{c} 
	=0,
\end{equation}
\begin{equation}
	\label{eq34}
	\left| E \right|\varphi_{c} =\sqrt {m^{2}+{\rm {\bf p}}^{2}} \varphi 
	_{c} .
\end{equation}
	\item[(ii)] $A^{0} \left( {{\rm {\bf x}}} \right)=0,\,\,\,A^{i} \left( {{\rm {\bf x}}} \right)\ne 0$ is the case of the constant magnetic field.

Here, the operator $A_{\left( + \right)} $ in (\ref{eq32}) also commutes with the 
summands on the left and
\begin{equation}
	\label{eq35}
	\left( {\left| E \right|^{2}-m^{2}-\left( {{\rm {\bm {\sigma \pi} }}} 
		\right)^{2}} \right)\varphi_{c} =0,
\end{equation}
\begin{equation}
	\label{eq36}
	\left| E \right|\varphi_{c} =\sqrt {m^{2}+\left( {{\rm {\bf p}}-eA} 
		\right)^{2}+e{\rm {\bm \sigma H}}} \,\varphi_{c} .
\end{equation}
Here, ${\rm {\bf H}}$ is the constant magnetic field.
	\item[(iii)] $A^{0} \left( {{\rm {\bf x}}} \right)\ne 0,\,\,\,A^{i} \left( {{\rm {\bf x}}} \right)=0$.
	
In this case, the operator $A_{\left( + \right)} $ in (\ref{eq32}) does not 
commute with the summands on the left, therefore we should consider Eq. 
(\ref{eq29}) with the wave function in the FGM representation
\begin{equation}
	\label{eq37}
	\left( {\left( {\left| E \right|-eA^{0}} \right)^{2}-m^{2}-{\rm {\bf 
				p}}^{2}+ie{\rm {\bm \sigma }}\nabla A^{0}} \right)\varphi_{FG} \left( 
	{{\rm {\bf x}}} \right)=0.
\end{equation}
Provided condition (\ref{eq26}) is fulfilled, from Eq. (\ref{eq21}), taking into 
account Eq. (\ref{eq23}), the Hamiltonian $H_{FW}^{\left( + \right)} $ is
\begin{equation}
	\label{eq38}
	\begin{array}{l}
	H_{FW}^{\left( + \right)} =H_{even}^{\left( + \right)} =A_{\left( + \right)} 
	\left[ {{\bm {\sigma \pi }}+eA^{0}+\dfrac{2m^{2}}{\left| E \right|+{\bm 
				{\sigma \pi }}-eA^{0}}}-  \right.   \\ [10pt]
		\left. 	- \dfrac{m^{2}}{\left| E \right|+{\bm {\sigma\pi}}-eA^{0}}\left( {{\bm {\sigma \pi }}-eA^{0}} \right)\dfrac{1}{\left| 
			E \right|+{\bm {\sigma \pi }}-eA^{0}} \right]A_{\left( + \right)} .
	\end{array}
\end{equation}
Here, $\left| E \right|$ is determined by the solution of Eqs. 
(\ref{eq29}) or (\ref{eq32}). The absence of the commutation $A_{\left( + \right)} $ with the first two summands in (\ref{eq38}) does not allow simplifying the expression for $H_{FW}^{\left( + \right)} $.
\end{itemize}
\subsection{Hamiltonians and FW energy operators for negative energies
	$\varepsilon =-\left| E \right|<0$}
\label{Subsec3.2}

In this case, by analogy with Eqs. (\ref{eq25}) - (\ref{eq32}),
\begin{equation}
	\label{eq39}
	\begin{array}{l}
		H_{FW}^{\left( - \right)} =U_{FW}^{\left( - \right)} H_{D} 
		U_{FW}^{\left( - \right) \dag}=H_{even}^{\left( - \right)} 
		+H_{odd}^{\left( - \right)} ,\, \\ [10pt]
		H_{odd}^{\left( - \right)} \,\psi_{FW}^{\left( - \right)} =0, \\ 
	\end{array}
\end{equation}
\begin{equation}
	\label{eq40}
	\begin{array}{l}
		H_{odd}^{\left( - \right)} \,\psi_{FW}^{\left( - \right)} =A_{\left( - 
			\right)} \beta_{c} m\left[ {1-\dfrac{m^{2}}{\left( {\left| E \right|+{\rm {\bm {\sigma \pi} }}+eA^{0}} \right)^{2}}} \right.-\dfrac{1}{\left| E \right|+{\rm 
				{\bm {\sigma \pi} }}+eA^{0}}\left( {{\rm {\bm {\sigma \pi} }}+\gamma^{5}_{c} 
			eA^{0}} \right)- \\ [10pt]
		\left. {-\left( {{\rm {\bm {\sigma \pi} }}-\gamma^{5}_{c} eA^{0}} 
			\right)\dfrac{1}{\left| E \right|+{\rm {\bm {\sigma \pi} }}+eA^{0}}} 
		\right]A_{\left( - \right)} \psi_{FW}^{\left( - \right)} =0. \\ 
	\end{array}
\end{equation}
Taking into account Eq. (\ref{eq24}), the numerator and denominator of the last summand in (\ref{eq40}) mutually commute.

Then,
\begin{equation}
	\label{eq41}
	H_{odd}^{\left( - \right)} \,\psi_{FW}^{\left( - \right)} =A_{\left( - 
		\right)} \beta_{c} m\left[ {1-\frac{m^{2}}{\left( {\left| E \right|+{\rm {\bm 
						{\sigma \pi} }}+eA^{0}} \right)^{2}}-\frac{1}{\left| E \right|+{\rm {\bm 
					{\sigma \pi} }}+eA^{0}}2{\rm {\bm 
				{\sigma \pi} }}} \right]A_{\left( - \right)} 
	\psi_{FW}^{\left( - \right)} =0.
\end{equation}
Equation (\ref{eq41}) can be written either in the FGM representation or in the FW representation
\begin{equation}
	\label{eq42}
	\left( {\left( {\left| E \right|+eA^{0}} \right)^{2}-\left( {{\rm {\bm 
					{\sigma \pi}}}} \right)^{2}-m^{2}-ie{\rm {\bm \sigma }}\nabla A^{0}} 
	\right)\chi_{FG} \left( {{\rm {\bf x}}} \right)=0,
\end{equation}
\begin{equation}
	\label{eq43}
	\chi_{FG} \left( {{\rm {\bf x}}} \right)=A_{\left( - \right)} \chi 
	_{c} \left( {{\rm {\bf x}}} \right),
\end{equation}
\begin{equation}
	\label{eq44}
	\left[ {A_{\left( - \right)}^{-1} \left( {\left( {\left| E \right|+eA^{0}} 
			\right)^{2}-\left( {{\rm {\bm 
						{\sigma \pi} }}} \right)^{2}-m^{2}-ie{\rm {\bm 
					\sigma }}\nabla A^{0}} \right)A_{\left( - \right)} } \right]\chi_{c} 
	\left( {{\rm {\bf x}}} \right)=0.
\end{equation}
The Hamiltonian $H_{FW}^{\left( - \right)} $ is
\begin{equation}
	\label{eq45}
	\begin{array}{l}
		H_{FW}^{\left( - \right)} =H_{even}^{\left( - \right)} =A_{\left( - \right)} 
	\left[ {-{\rm {\bm {\sigma \pi} }}+eA^{0}-\dfrac{2m^{2}}{\left| E \right|+{\rm {\bm {\sigma \pi} }}+eA^{0}}}+ \right. \\ [10pt]
	\left. 	\dfrac{m^{2}}{\left| E \right|+{\rm {\bm 
					{\sigma \pi} }}+eA^{0}}\left( {{\rm {\bm 
					{\sigma \pi} }}+eA^{0}} \right)\dfrac{1}{\left| 
			E \right|+{\rm {\bm 
					{\sigma \pi} }}+eA^{0}} \right]A_{\left( - \right)} .
	\end{array}
\end{equation}
In case of the free motion or the motion in the constant magnetic field, 
Eqs. (\ref{eq33}) - (\ref{eq36}) are valid for negative energies $\varepsilon <0$ with the substitution of $\varphi_{c} \left( {{\rm {\bf x}}} \right)$ by 
$\chi_{c} \left( {{\rm {\bf x}}} \right)$.

If $A^{0}\left( {{\rm {\bf x}}} \right)\ne 0,\,\,\,A^{i}\left( {{\rm {\bf 
			x}}} \right)=0$, then, for $\varepsilon <0$, equations (\ref{eq42}) - (\ref{eq45}) are valid 
with the substitution of ${\rm {\bm 
		{\sigma \pi} }}\to {\rm {\bm \sigma \bf p}}$.


\section{The FGM representation and its relation with the FW representation}
\label{Sec4}

Let us write Dirac equation (\ref{eq1}) as
\begin{equation}
\label{eq46}
\left[ {\left( {p^{0}-eA^{0}} \right)-{\rm {\bm { \alpha \pi} }}-\beta m} 
\right]\psi_{D} ({\rm {\bf x}},t)=0.
\end{equation}
Dirac (see, for example, Ref. \cite{bib12}) multiplied Eq. 
(\ref{eq46}) on the left by the multiplier $\left( {p^{0}-eA^{0}} \right)+{\rm {\bm 
	{\alpha \pi} }}+\beta m$ and obtained the equation of the second order
\begin{equation}
\label{eq47}
\left[ {\left( {p^{0}-eA^{0}} \right)^{2}-\left( {{\rm {\bf p}}-e{\rm {\bf 
				A}}} \right)^{2}-m^{2}+e{\rm {\bf \Sigma H}}-ie{\rm {\bm \alpha \bf E}}} 
\right]\psi ({\rm {\bf x}},t)=0.
\end{equation}
In Eq. (\ref{eq47}), $\psi ({\rm {\bf x}},t)$ is the bispinor wave 
function; $p^{0}=i\dfrac{\partial }{\partial t};\,\,\,{\rm {\bf 
	p}}=-i\vec{{\nabla }};$ $A^{0}\left( {{\rm {\bf r}},t} 
\right)$ and $A^{i}\left( {{\rm {\bf r}},t} \right)$ are electromagnetic 
potentials; ${\rm {\bf H}}=\rm rot{\rm {\bf A}},\,\,{\rm {\bf 
	E}}=-\dfrac{\partial {\rm {\bf A}}}{\partial t}-\nabla A^{0}$ are magnetic and 
electric fields; ${\rm {\bf \Sigma }}=\left( {{\begin{array}{*{20}c}
		{{\rm {\bm \sigma }}} \hfill & 0 \hfill \\
		0 \hfill & {{\rm {\bm \sigma }}} \hfill \\
\end{array} }} \right)$.

In case of static fields, we will consider stationary states when $p^{0}\psi 
=\varepsilon \psi $.

For Dirac matrices in the chiral representation, $\alpha^{i}_{c}=\left( 
{{\begin{array}{*{20}c}
		{\sigma^{i}} \hfill & \,\,\,\, 0 \hfill \\
		0 \hfill & {-\sigma^{i}} \hfill \\
\end{array} }} \right)$ and in this case, in Eq. (\ref{eq47}) there is no 
mixing of the upper and lower components of a bispinor $\psi_{FG} $. If
\begin{equation}
\label{eq48}
\psi_{FG} \left( {{\rm {\bf x}},t} \right)=S \psi  \left( {{\rm {\bf x}},t} \right) = \left( {{\begin{array}{*{20}c}
			{\varphi_{FG} \left( {{\rm {\bf x}}} \right)} \hfill \\
			{\chi_{FG} \left( {{\rm {\bf x}}} \right)} \hfill \\
\end{array} }} \right)e^{-i\varepsilon t},
\,\,\,\,\,\,\,S=\dfrac{1}{\sqrt 2} 
\left( {{\begin{array}{*{20}c}
			{{ I }} \hfill & \,\,\, \,I \hfill \\
			I \hfill & {{ -I }} \hfill \\
\end{array} }} \right),
\end{equation}
then Eq. (\ref{eq47}) is reduced to the two separate equations for the spinors 
$\varphi_{FG} \left( {{\rm {\bf x}}} \right),\,\,\,\chi_{FG} \left( {{\rm 
	{\bf x}}} \right)\,\,$.
\begin{equation}
\label{eq49}
\left[ {\left( {\varepsilon -eA^{0}} \right)^{2}-\left( {{\rm {\bf p}}-e{\rm 
			{\bf A}}} \right)^{2}-m^{2}+e{\rm {\bm \sigma \bf H}}-ie{\rm {\bm \sigma \bf E}}} 
\right]\varphi_{FG} ({\rm {\bf x}})=0,
\end{equation}
\begin{equation}
\label{eq50}
\left[ {\left( {\varepsilon -eA^{0}} \right)^{2}-\left( {{\rm {\bf p}}-e{\rm 
			{\bf A}}} \right)^{2}-m^{2}+e{\rm {\bm \sigma \bf H}}+ie{\rm {\bm \sigma \bf E}}} 
\right]\chi_{FG} ({\rm {\bf x}})=0.
\end{equation}
Earlier, similar equations were considered by Feynman and Gell-Mann \cite{bib8}.

It is particularly remarkable that Eqs. (\ref{eq49}), (\ref{eq50}) are connected with the equations in the FW representation (see (\ref{eq29}) - (\ref{eq32}) and (\ref{eq42}) -(\ref{eq44})). With such a connection, the problem of ''extraneous'' solutions in 
equations (\ref{eq47}), (\ref{eq49}) and (\ref{eq50}) is solved automatically.

Equation (\ref{eq49}) must be used for positive energies $\varepsilon =\left| E 
\right|>0$. In this case,
\begin{equation}
\label{eq51}
\varphi_{FG} \left( {{\rm {\bf x}}} \right)=A_{\left( + \right)} \,\varphi 
_{c} \left( {{\rm {\bf x}}} \right).
\end{equation}
Equation (\ref{eq50}) must be used for negative energies $\varepsilon =-\left| E 
\right|<0$.

In this case,
\begin{equation}
\label{eq52}
\chi_{FG} \left( {{\rm {\bf x}}} \right)=A_{\left( - \right)} \chi_{c} 
\left( {{\rm {\bf x}}} \right).
\end{equation}
For the case of a centrally symmetric field $eA^{0}\left( r 
\right)$ and $A^{i}\left( {{\rm {\bf x}}} \right)=0$ (see App. A), it is shown 
that the energy spectra of the FGM equations contain the 
spectrum of the initial Dirac equation.


\section{Conclusions}
\label{Conclusions}

By using Dirac matrices in the chiral representation, the closed expressions 
for energy operators in the FW representation in the presence 
of electrostatic fields were first obtained. The obtained expressions can be 
used in the nonperturbative quantum electrodynamics with strong electric 
fields.

In this paper, the connection was first established between the 
FW and FGM representations.


\section*{Acknowledgments}

This study was conducted within the framework of the scientific program of 
the National Center for Physics and Mathematics, section N5 ''Particle 
physics and cosmology. Stage 2023-2025''.

The author thank A.L.Novoselova for the essential technical assistance in 
the preparation of the paper.


\section*{Appendix A. Separation of variables in the FGM equations with a centrally symmetric Coulomb field}

The FGM equations  (see (\ref{eq49}) and (\ref{eq50})) were obtained in the Sec. 4 with the use 
of the Dirac matrices in the chiral representation.

For the case $A^{i}\left( {{\rm {\bf x}}} \right)=0$, the equations FG in the 
Coulomb field $eA^{0}\left( r \right)$ have the form

$$\left[ {\left( {\varepsilon -eA^{0}} \right)^{2}-{\rm {\bf 
				p}}^{2}-m^{2}-ie{\rm {\bm \sigma \bf E}}} \right]\varphi_{FG} ({\rm {\bf 
			x}})=0, \eqno(\rm {A.1})
	\label{app1}
$$

$$	\left[ {\left( {\varepsilon -eA^{0}} \right)^{2}-{\rm {\bf 
				p}}^{2}-m^{2}+ie{\rm {\bm \sigma \bf E}}} \right]\chi_{FG} ({\rm {\bf x}})=0.
	\label{app2} \eqno(\rm {A.2})
$$

The chiral representation (\ref{eq3}) is connected with the 
standard representation (\ref{eq2}) by the unitary transformation

$$S=\dfrac{1}{\sqrt 2 }\left( {{\begin{array}{*{20}c}
				I \hfill & \,\,\,\,I \hfill \\
				I \hfill & {-I} \hfill \\
	\end{array} }} \right).
   \label{app3} \eqno(\rm {A.3})
$$

The initial Dirac equation (\ref{eq1}) and (\ref{eq5}) with $A^{i}\left( {{\rm {\bf x}}} 
\right)=0$ and with representation (\ref{eq2}) allows the separation of variables, if, 
in the spherical coordinate system, a bispinor $\psi_{D} \left( {{\rm {\bf 
			r}},t} \right)=\psi_{D} \left( {r,\theta ,\varphi ,t} \right)$ is defined as

$$\psi_{D} \left( {r,\theta ,\varphi ,t} \right)=\left( 
	{{\begin{array}{*{20}c}
				\,\,\,\,\,\,{F\left( r \right)\xi \left( \theta \right)} \hfill \\ [5pt]
				{-iG\left( r \right)\sigma^{3}\xi \left( \theta \right)} \hfill \\
	\end{array} }} \right)e^{-i\varepsilon t}e^{im_{\varphi } \varphi }.
\label{app4} \eqno(\rm {A.4})
$$

As a result, we obtain the system of equations for the real radial 
functions $F\left( r \right),\,\,G\left( r \right)$

$$ \begin{array}{l}
	\dfrac{dF}{dr}+\dfrac{1+\kappa }{r}F-\left( {\varepsilon -eA^{0}+m} 
	\right)G=0, \\ [10pt]
	\dfrac{dG}{dr}+\dfrac{1-\kappa }{r}G+\left( {\varepsilon -eA^{0}-m} 
	\right)F=0. \\ 
\end{array}
	\label{app5} \eqno(\rm {A.5})
$$

In (A.4), $\xi \left( \theta \right)$ is a spherical harmonic for the 
half-spin, $m_{\varphi } =-j,-j+1,\,...\,j$ is a projection of total moment, 
$\kappa =\mp 1,\mp 2...=\left\{ {\begin{array}{l}
		-\left( {l+1} \right),\,\,j=l+1 / 2 \\ 
		\,\,\,\,\,\,\,\,l,\,\,\,\,\,\,\,\,\,\,\,\,\,\,\,\,j=l-1 / 2 \\ 
\end{array}} \right.$; $j,l$ are the quantum numbers of the total and 
orbital moments of the Dirac particle.

The transition to the Dirac equation with the chiral representation (\ref{eq3}) is 
implemented by the unitary transformation (A.3). The transformed wave 
function is

$$ \psi_{c} \left( {r,\theta ,\varphi ,t} \right)= S\psi_{D} \left( {r,\theta 
		,\varphi ,t} \right)= \frac{1}{\sqrt{2}}\left( {{\begin{array}{*{20}c}
				{\left( {F\left( r \right)-iG\left( r \right)\sigma^{3}} \right)\xi \left( 
					\theta \right)} \hfill \\ [5pt]
				{\left( {F\left( r \right)+iG\left( r \right)\sigma^{3}} \right)\xi \left( 
					\theta \right)} \hfill \\
	\end{array} }} \right)e^{-i\varepsilon t}e^{im_{\varphi } \varphi }.
\label{app6} \eqno(\rm {A.6})
$$

After separation of variable, we again arrive at the system of equations for the radial functions (A.5).

Equation (A.6) shows, that in the FGM equations (A.1) and (A.2)

$$	\varphi_{FG} \left( {r,\theta ,\varphi ,t} \right)=\frac{1}{\sqrt{2}} \left( {F\left( r 
		\right)-iG\left( r \right)\sigma^{3}} \right)\xi \left( \theta 
	\right)e^{-i\varepsilon t}e^{im_{\varphi } \varphi },
	\label{app7} \eqno(\rm {A.7})
$$

$$	\chi_{FG} \left( {r,\theta ,\varphi ,t} \right)=\frac{1}{\sqrt{2}} \left( {F\left( r 
		\right)+iG\left( r \right)\sigma^{3}} \right)\xi \left( \theta 
	\right)e^{-i\varepsilon t}e^{im_{\varphi } \varphi }.
	\label{app8} \eqno(\rm {A.8})
$$

Substituting (A.7) and (A.8), respectively, in Eqs. (A.1) and (A.2) and 
replacing for the centrally symmetric electrostatic field ${\rm {\bm \sigma 
		\bf E}}\to \sigma^{3}\dfrac{dA^{0}}{dr}$, we obtain the same system of the real 
equations for the radial functions $F\left( r \right),\,\,G\left( r 
\right)$\footnote{At the separation of variables with the use of functions $\xi 
	\left( \theta \right)$ in the initial Dirac equation, the equivalent 
	replacement of matrices $\alpha^{1}\to \alpha^{3},\,\,\,\alpha^{3}\to 
	\alpha^{2},\,\,\,\alpha^{2}\to \alpha^{1}$ is carried out. As a result, in 
	the spherical coordinate system, the index 3 corresponds to the direction along 
	$r$ for the Dirac and Pauli matrices.}.

$$ \begin{array}{l}
	\left( {\left( {\varepsilon -eA^{0}} \right)^{2}-m^{2}-\dfrac{\kappa \left( 
			{\kappa +1} \right)}{r^{2}}+\dfrac{d^{2}}{dr^{2}}+\dfrac{2}{r}\dfrac{d}{dr}} 
	\right)F\left( r \right)=-e\dfrac{dA^{0}}{dr}G\left( r \right), \\ [10pt]
	\left( {\left( {\varepsilon -eA^{0}} \right)^{2}-m^{2}-\dfrac{\kappa \left( 
			{\kappa +1} \right)}{r^{2}}+\dfrac{d^{2}}{dr^{2}}+\dfrac{2}{r}\dfrac{d}{dr}} 
	\right)G\left( r \right)=e\dfrac{dA^{0}}{dr}F\left( r \right). \\ 
\end{array}
\label{app9} \eqno(\rm {A.9})
$$

The substitution the equations for the Dirac wave functions (A.5) in (A.9) 
shows their mutual consistency.

Thus, an energy spectrum of the FGM equations contains an 
energy spectrum of the Dirac equation.


\end{document}